\pdfoutput=1
		\documentclass{isprs}
		\usepackage{hyperref}
		\usepackage{subfigure} 
		\usepackage[utf8]{inputenc}
	
		\usepackage{varioref} 
		\usepackage{natbib} % necessary to control spacing
	\usepackage{enumitem}
		\usepackage{graphicx}
		\DeclareGraphicsExtensions{.pdf,.jpg,.png}

		\usepackage[usenames,dvipsnames]{color}
	\usepackage{amsmath}
		\usepackage{algorithm2e}
		\usepackage{microtype}  
		\usepackage{SIunits} 
		\usepackage{listings}
		\usepackage{tabularx} 
		\usepackage{setspace}

		\newcommand{\myimage}[3]
					{
					\begin{figure} [h!]
						\begin{center}
							\includegraphics[width=\linewidth,keepaspectratio]{#1}
							\caption{#2}  
							\label{#3}
							\end{center}
					\end{figure} 
					}

		\newcommand{\myimageHL}[4]
		{
			\begin{figure} [ht!]
				\begin{center}
					\includegraphics[width= #4 \linewidth ,keepaspectratio]{#1}
					\caption{#2}  
					\label{#3}
				\end{center}
			\end{figure} 
		}

		\newcommand{\myimageFullPageWidth}[3]
								{
								\begin{figure*}[ht]
									\begin{center}
										\includegraphics[width=\textwidth,keepaspectratio ]{#1}
										\caption{#2}  
										\label{#3} 
										\end{center}
								\end{figure*} 
								}

	\usepackage{rotating}

\title{StreetGen : In base city scale procedural generation of streets: road network, road surface and street objects}
\author{Rémi Cura  $^{A}$, Julien Perret $^A$, Nicolas Paparoditis  $^A$}
\address{ $^A$  Université Paris-Est, IGN, SRIG, COGIT \& MATIS, 73 avenue de Paris, 94160 Saint Mandé, France\\
	first\_name.last\_name@ign.fr
	}

\begin{document}
 
%\begin{keyword}
%RDBMS, point cloud, filtering, indexing, compression, point cloud storage, point cloud I/O, point cloud generalisation, patch, point grouping, point cloud visualisation
%\end{keyword}

%% Inserting the abstract
%% ---------------------------------------------------------------------
%% Copyright 2014, Thales, IGN, Rémi Cura
%% 
%% This file is the abstract of the article
%% ---------------------------------------------------------------------

%\section{Abstract}     

\abstract{ 	
%Streets are the basis of cities, and place of great importance for the Billions of people that live in it.
Streets are large, diverse, and used for several (and possibly conflicting) transport modalities as well as social and cultural activities. Proper planning is essential and requires data.
Manually fabricating data that represent streets (street reconstruction) is error-prone and time consuming.
Automatising street reconstruction is a challenge because of the diversity, size, and scale of the details ($\sim \centi \meter$ for cornerstone) required.
The state-of-the-art focuses on roads (no context, no urban features) and is strongly determined by each application (simulation, visualisation, planning).
We propose a unified framework that works on real Geographic Information System (GIS) data and uses a strong, yet simple modelling hypothesis when possible to robustly model streets  at the city level or street level.
Our method produces a coherent street-network model containing topological traffic information, road surface and street objects.
We demonstrate the robustness and genericity of our method by reconstructing the entire city of Paris streets and exploring other similar reconstruction (airport driveway).
}

\maketitle 
%\tableofcontents

%\myimageFullPageWidth{./illustrations/chap2/lod_banner/banner_for_paper}{Graphical Abstract : a Lidar point cloud (1), is split it into patches (2) and stored in a Point Cloud Server, patches are re-ordered to obtain free LOD (3) (a gradient of LOD here), lastly the ordering by-product is a multiscale dimensionality descriptor used as a feature for learning and efficient filtering (4).}{lod.fig:banner_image}
\myimageFullPageWidth{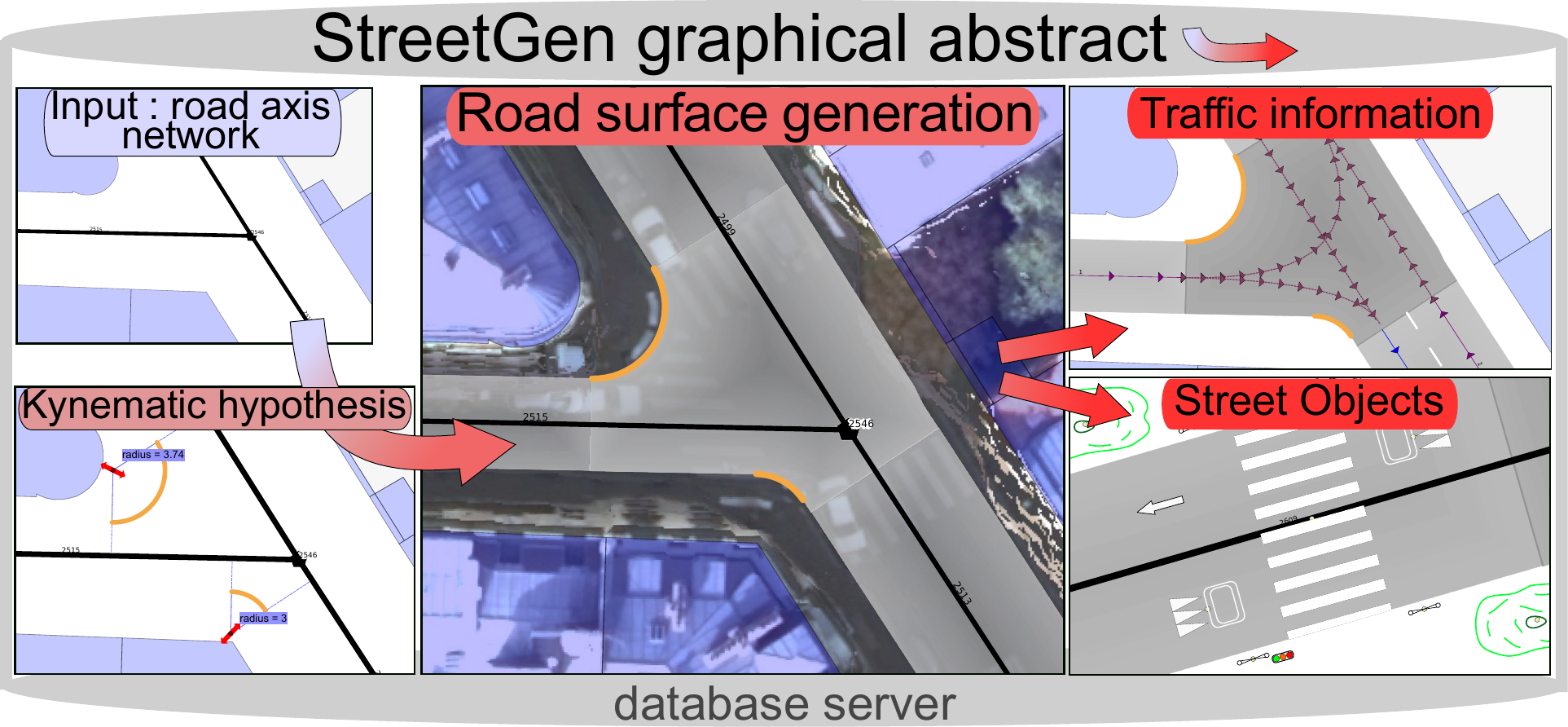}{StreetGen graphical Abstract. }{sg.fig.graphical_abstract}

%% Inserting the introduction
	%% ---------------------------------------------------------------------
%% Copyright 2014, Thales, IGN, Rémi Cura
%% 
%% This file contains the introduction of article
%% ---------------------------------------------------------------------

\section{Introduction}
\myimageFullPageWidth{"illustrations/chap3/streetgen/bandeau/bandeau"}{StreetGen in a glance. Given road axes, reconstruct network, find corner arcs, compute surfaces, add lanes and markings.}{sg.fig:header}
%\subsection{Problem} 
	%More than half of the World population live in urban areas.
	%At such, cities are essential places of interchange.
	%Those contacts happen in the common medium: the streets. 
	Streets are complex and serve many types of purposes, including practical (walking, shopping, etc.), social (meeting, etc.), and cultural (art, public events, etc.).
	%Concentrating many people in those streets raise issues, and thus those places must be managed carefully, be it for planing, understanding, or reworking.
	Managing existing streets and planning new ones necessitates data, as planning typically occurs on an entire neighbourhood scale.
	These data can be fabricated manually (cadastral data, for instance, usually are).
    Unfortunately, doing so requires immense resources in time and people.
    
	Indeed, a medium sized city may have hundreds of kilometers of streets.
	Streets are not only spatially wide, they also are very plastic and change frequently.
	Furthermore, street data must be precise because some of the structuring elements, like cornerstones (they separate sidewalks from roadways) are only a few centimetres in height.
	Curved streets are also not adapted to the Manhattan hypothesis, which states that city are organised along three dominant orthogonal directions \cite{Coughlan1999}.
	
	The number and diversity of objects in streets are also particularly challenging.
	Because street data may be used for very different purposes (planning, public works, and transport design), it should be accessible and extensible. 
 
%\subsection{State Of The Art}
%	The state of the art mainly focus on reconstructing road, and not street.
%	Moreover, the focus is often in semi-urban area, where the road path is more logical (and mathematical) than in cities.
	 
	Traditionally, street reconstruction solutions are more road reconstruction and are also largely oriented by the subsequent use of the reconstructed data.
	For instance, when the use is traffic simulation~\cite{Wilkie2012,Nguyen2014,Yeh2015}, the focus is on reconstructing the road axis (sometime lanes), not necessarily the roadway surface.
	In this application, it is also essential that the reconstructed data is a network (with topological properties) because traffic simulation tools rely on it.
	However, the focus is clearly to reconstruct road and not streets. Streets are much more complex objects than roads, as they express the complexity of a city, and contains urban objects, places, temporary structures (like a marketplace).
    The precision of the reconstruction is, at best, around a metre in terms of accuracy. 
	
	Another application is road construction for the virtual worlds or driving simulations.
	In this case, we may simply want to create realistic looking roads. For this, it is possible to use real-life civil engineering rules, for instance using a clothoid as the main curve in highway \cite{McCrae2009a,Wang2014}.
	When trying to produce a virtual world, the constructed road must blend well into its environment. For instance, the road should pass on a bridge when surrounded by water. We can also imitate real-world road-building constraints, and chose a path for the road that will minimise costs~\cite{Galin2010}.
	Roads can even be created to form a hierarchical network~\cite{Galin2011}.
	Such generated roads are nice looking and blend well into the terrain, but they do not match reality. That is, they only exist in the virtual world.
	
	The aim may also be to create a road network as the base layout of a city.
	Indeed, stemming from the seminal work of~\cite{Parish2001}, a whole family of methods first creates a road network procedurally, then creates parcels and extrudes these to create a virtual city.
	These methods are very powerful and expressive, but they may be difficult to control (that is, to adapt the method to get the desired result). Other works focus on control method ~\cite{Chen2008,Lipp2011,Benes2014}.
	Those methods suffer from the same drawback; they are not directly adapted to model reality.
	
    More generally, given procedural generation methods, finding the parameters so that the generated model will match the desired result is still an on-going issue (inverse procedural modelling, like in \cite{Martinovic2013} for façade, for instance).

%	In this article we try to mix the advantages of several approaches. 
	In this work, we propose an original approach to the procedural modelling of streets : StreetGen.
	We start from rough GIS data (Paris road axis); thus, our modelling is based on a real road network.
	Then, we use a basic hypothesis and a simple road model to generate more detailed data.
	At this point, we generate street data for a large city (Paris); the result is one street network model.
	We use a widespread street network model, where the skeleton is formed by street axis and intersection forming a network. Then other constituents (lane, pedestrian crossing, markings, etc.) are linked to this network.
	We base all our work on a Relational DataBase Management System (RDBMS), to store inputs, results, topology, and processing methods.

%\subsection{Plan}
	
	In Section~\ref{sg.sec:method} we explain why we chose to base our work on a RDBMS, and explain the hypothesis, how we generate the road surface, and how the parameters of the resulting model can be edited.
	In Section~\ref{sg.sec:result} we provide results of street generation and results of editing.
	In Section~\ref{sg.sec:discussion}, we discuss the results and present limitations and possible improvements.

%	\newpage

%% Inserting the Method part
	%% ---------------------------------------------------------------------
%% Copyright 2014, Thales, IGN, Rémi Cura
%% 
%% This file present the method of the article
%% ---------------------------------------------------------------------

\section{Method}
	\label{sg.sec:method}
\subsection{Introduction to StreetGen} 
	%\includepdf[]{"illustrations/chap3/overall/overall"}
	\myimage{"illustrations/chap3/overall/overall"}{StreetGen workflow.}{sg.fig:overall}
	
	The design of StreetGen is a result of a compromise between theoretical and practical considerations. StreetGen amplifies data using a simple, yet strong hypothesis. 
	As such, the approach is to attain correct results when the hypothesis appears correct and change the method to something more robust when the hypothesis appears wrong, so as to always have a best guess result.
	
	Second, StreetGen has been designed to work independently at different level.
	It can generates street data at the city level.
	The exact same method also generates street data interactively at the street level.
	
	Lastly, StreetGen results are used by different applications (visualisation, traffic simulation, and spatial analysis). As such, the result is a coherent street data model with enforced constraints, and we also keep links with input data (traceability). 
	
	Figure \ref{sg.fig:overall} sum up StreetGen process.

\subsection{Introduction to RDBMS}
	
	We chose to use a RDBMS (*\cite{PostgreSQL2014} with \cite{PostGIS2014}) at the heart of StreetGen for many reasons.
	First, RDBMSs are classical and widespread, which means that any application using our results can easily access it, whatever the Operating System (OS) or programming language.
	Second, RDBMSs are very versatile and, in one common framework, can regroup our input GIS data, a road network (with topology), the resulting model of streets, and even the methods to create it.
	Unlike file-based solutions, we put all the data in relation and enforce these relations. For instance, our model contains surfaces of streets that are associated with the corresponding street axis. If one axis is deleted, the corresponding surface is automatically deleted.
	We push this concept one step further, and link result tables with input tables, so that any change in input data automatically results in updating the result.
	Lastly, using RDBMS offers a multi OS, multi GIS (many clients possible), multi user capabilities, and has been proven to scale easily.
	We stress that the entirety of StreetGen is self contained into the RDBMS (input data, processing methods, and results).

\subsection{StreetGen Design Principles}
 	\paragraph{Input of StreetGen}
 	We use few input data, and accept that these are fuzzy and may contain errors.\\
 	The first input is a road axis network made of polylines with an estimated roadway width for each axis. We use the BDTopo\footnote{\url{http://professionnels.ign.fr/bdtopo}} product for Paris in our experiment, but this kind of data is available in many countries.
 	It can also be reconstructed from aerial images \cite{Montoya-Zegarra2014}, Lidar data \cite{Poullis2010}, or tracking data (GPS and/or cell phone) \cite{Ahmed2014}.\\
 	Using the road axis network, we reconstruct the topology of the network up to a tolerance using either GRASS GIS (\cite{neteler2012grass}) or directly using PostGIS Topology \cite{PostGISTopology2014}.
 	We store and use the network with valid topology with PostGIS Topology.
	\\
	The second input is the roughly estimated average speed of each axis. We can simply derive it from road importance, or from road width (when a road is wide, it is more likely that the average speed will be higher).
	\\
 	The third input is our modelling of streets and the hypothesis we create.
 	
 	Because the data we need can be reconstructed and there is a low requirement on data quality, our method could be used almost anywhere.
 	In particular, road attributes may be very basic and can still be corrected if necessary.
 	
 	\paragraph{Street data model} 
 	\myimageFullPageWidth{"illustrations/chap3/streetgen/street_data_model/street_data_model"}{Street data model.}{sg.fig:street_data_model}
 	Real life streets are extremely complex and diverse; we do not aim at modelling all the possible streets in all their subtleties, but rather aim at modelling typical streets with a reasonable number of parameters.
 	
 	First, we observe that street and urban objects are structured by the street axis.
 	For instance, a pedestrian crossing is defined with respect to the street axis.
 	At such, we centre our model on street axes.

 	Second, we observe that streets can be divided into two types: parts that are morphologically constant (same roadway width, same number of sidewalks, etc.), and transition parts (intersection, transition when the roadway width increases or decreases).\\
 	We follow this division so that our street model is made of morphologically constant parts (section) and transition parts (intersection).
 	The separation between transition and constant parts is the section limit and is expressed regarding the street axis in curvilinear abscissa.
 	
 	Third, classical streets are adapted to traffic, which means that a typical vehicle can safely drive along the street at a given speed.
 	This means that cornerstone in an intersection does not form sharp right turns that would be dangerous for vehicle tires. The most widespread cornerstone path in this case seems to be the arc of a circle, as it is the easiest form to build during public work.
 	Therefore, we consider cornerstone path to be either a segment or the arc of a circle.
 	This choice is similar to \cite{Wilkie2012} and is well adapted to the city,
 	but not so well adapted to peri-urban roads, where the curve of choice is usually the clothoid (like in \cite{McCrae2009}), because it is actually the curve used to build highways and fast roads.
 	
 	The surface of intersection is then defined by the farthest points on each axis where the border curve starts.
 	In this base model, we add lanes, markings, etc.
 	  
	\paragraph{Kinematic rule of thumb}
	\label{sg.method.kinematic_hypothesis}
	\myimageHL{"illustrations/chap3/streetgen/radius_type/radius_type"}{3 different radius size (3m, 4.9m, 7.6 m) for streets of various importancy, from real Paris data}{sg.fig:radius_type}{1}
	
	We propose basic hypotheses to attempt to estimate the radius of the corner in the intersection.
	We emphasise that these are rules of thumb that give a reasonable best guess result, and does not mean that the streets were actually made following these rules (which is false for Paris for instance).
	
	Our first hypothesis is that streets were adapted so that vehicles can drive conveniently at a given speed $s$ that depends on the street type.
	For instance, vehicles tend to drive more slowly on narrow residential streets than on city fast lanes.\\
	Our second hypothesis is that given a speed, a vehicle is limited in the turns it can make. Considering that the vehicle follows an arc of circle trajectory, a radius that is too small would produce a dangerous acceleration and would be uncomfortable. Therefore we are able to find the radius $r$ associated with a driving speed $s$ through an empirical function $f(s)->r$. This function is based on real observations of the French organisation SETRA \citep{Setra2006} (For function, see Section \ref{sg.eq.setra}).
	
	From our street data model and these kinematic rules of thumb, we deduce that if we roughly know the type of road, we may be able to roughly estimate the speed of the vehicles on it. From the speed, we can estimate a turning radius, which leads to the roadway geometry (See Figure \ref{sg.fig:radius_type}).
	
	Schematically, we consider that a road border is defined by a vehicle driving along it at a given speed, while making comfortable turns.

\subsection{Robust and Efficient Computing of Arcs}
	\paragraph{Goal}
	The hypotheses in the above section allow us to guess a turning radius from the road type.
	This turning radius is used to reconstruct the arcs of a circle that limits the junctions.
	The method must be robust because our hypotheses are just best guesses and are sometime completely wrong.

	Given two road axis ($a_1,a_2$) that are each polylines, and \emph{not} segments), having each an approximate width ($w_1,w_2$) and an approximate turning radius ($r = min(r_1,r_2)$, or another choosing rule), we want to find the centre of the arc of the circle that a driving vehicle would follow.
	\myimage{"illustrations/chap3/radius_direct_computing/radius_direct_computing"}{Finding the circle centre problem. Left classical problem, middle and right using real-world data.}{sg.fig:circle center problem}
	
	\paragraph{Method}
	\myimageHL{"illustrations/chap3/circle_center/circle_center"}{A method to robustly find circle centres using geometric operations. Buffer of axis is computed, then the intersection of outer ring of buffer returns a set of candidates points. Among the candidates, the one closest to the intersection centre are the final circle centre.}{sg.gif:circle_center}{1}
	Our first method was based on explicit computing, as in \cite{Wang2014}{, Figure 13}.
	However, this method is not robust, and has special cases (flat angle, zero degree angle, one road entirely contained in another), is intricately two-dimensional (2D), and, most importantly, cannot be used on poly-lines.
	Yet real-world data is precisely made of poly-lines, due to data specification or errors.
	
	We choose to use morphological and boolean operations to overcome these limitations.
	Our main operators are positive and negative buffers (formally, the Minkowski sum of the input with a disk of given size) as well as the surface intersection, union, etc.
	
	We are looking for the centre of the arc of the circle. Thus, by definition the centre could be all the places of distance of $d_1 = w_1+r$ from $a_1$ and distance of $d_2 = w_2+r$ from $a_2$.
	\\
	We translate this into geometrical operations:
	\begin{itemize}[noitemsep,topsep=0pt,parsep=0pt,partopsep=0pt]
		\item $buffer_i$, buffer of $a_i$ with $d_i$
		\item $inter$, the intersection of boundary of buffers, which is commonly a set of point but can also be a set of points and curve. All those place could be circle centre.
		\item $closest$, the point of $inter$ that is the closest to the junction centre. We must filter this among the candidates in order to keep only the one that makes the most sense, given our hypotheses.
	\end{itemize}
	  
	\paragraph{When hypothesis are wrong}
	
	In some cases $closest$ may be empty (when one road is geometrically contained in another considering their width for instance). In this case our method fails with no damages, as no arc is created.\\
	The radius may not be adapted to the local road network topology. This predominantly happens when the road axis is too short with respect to the proposed radius. In this case, we reduce the guessed radius to its maximal possible value by explicitly computing the maximum radius if possible.\\
	It also happens that the hypotheses regarding the radius are wrong, which creates obviously misplaced arcs. We chose a very simple option to estimate whether an arc is misplaced or not and simply use a threshold on the distance between the arc and the centre of the intersection. In this case, we set the radius to a minimum that corresponds to the Paris lane separator stone radius (0.15 \meter).

\subsection{Computing Surfaces from Arc Centres}
	\paragraph{Border points}
	\myimageHL{"./illustrations/chap3/streetgen/border_point/border_point"}{From circle centres, border points that limit the intersection are found by projection and filtering (farthest from intersection centre).}{sg.fig:border_point}{1}
	When centre of circles are found, we can compute the associated arcs and find intersection limit (See Fig. \ref{sg.fig:border_point}).
	We create the corresponding arc of circles by projecting the centres of the circle on both axis buffered by $w_i$.
	In fact, we do not use a projection, as a projection on a polyline may be ill-defined (for instance projecting on the closest segment may not work). Instead, we take the closest point.
	
	Similarly, we 'project' the circle centre onto the road axis. We call these projections candidate border points. 
	We have two or less border points per axis per intersection.
	According to our intersection surface model, we only keep one of the candidates per axis per intersection, choosing the candidate that is the farthest from the intersection centre. We define the distance from the intersection centre by using the curvilinear abscissa, which is necessary because, in some odd cases, the Euclidian distance may be misleading.

	\paragraph{Section and intersection surface}
	\myimageHL{"./illustrations/chap3/section/local_cutting"}{Creating the border line by cutting the section following a local estimation of the normal.}{sg.fig:local_cutting}{1}
	We compute the section surface by first creating border lines at the end of each section out of border points.
	The border lines are normal to a local straight approximation of the road axis.
	Then, we use these lines to cut the bufferised road axis to obtain the surfaces of road axis that are within the intersection.

	At this point, it would be possible to construct the intersection surface by linking border lines to arcs, passing by the buffered road axis when necessary.
	We found it too difficult to do it robustly because some of the previous results may be missing or slightly false due to bad input data, wrong hypotheses or a computing precision issue.
	
		\myimageHL{"./illustrations/chap3/build_area/build_area"}{From cutted axis surfaces and arcs, the function ST\_BuildArea build the maximum area possible.}{sg.fig.build_area}{0.95}
	We prefer a less specific method.
	We use the "ST\_BuildArea" function (See Fig. \ref{sg.fig.build_area}). Given a set of geometries, it breaks all the geometries into polylines, then creates largest possible surface from those polylines.
	We use it on cut roads and arcs. 
	
	\paragraph{Variable buffer}
	\myimageHL{"illustrations/chap3/streetgen/variable_buffer/variable_buffer"}{Variable buffer for robust roadway width transition.}{sg.fig:variable_buffer}{1}
	In the special case where the intersection is only a change of roadway width, the arc of the circle transition is less realistic than a linear transition.
	We use a variable buffer to do this robustly. It also offers the advantage to being able to control the three most classical transitions (symmetric, left, and right) and the transition length using only the street axis.
	
	We define the variable buffer as a buffer whose radius is defined at each vertex (i.e., points for linestring).
	The radius varies linearly between vertices.
	One easy, but inefficient solution to compute it is to build circles and isosceles trapezoids and then union the surface of these primitives.
	We use the easy version.

	\paragraph{Lane, markings, street objects}
	
	Based on the street section, we can build lanes and lane separation markings.
	To this end, we cannot simply translate the centre axis because axis are polylines (See Fig. \ref{sg.fig:no_translation}).
	Instead, a function similar to a buffer has to be used ("ST\_OffsetCurve").
	\myimage{"illustrations/chap3/lane/no_translation"}{Starting from center line (black), a translation would not create correct a lane (red). We must use the buffer (green).}{sg.fig:no_translation} 
	
	Our input data contains an estimation of the lane number. Even when such data is missing, it can still be guessed from road width, road average speed, etc., using heuristics. The number of lane could also be retrieved from various remote sensing data. For instance, \cite{Jin2009} propose to use aerial images.
	We can also build pedestrian crossings along the border lines.

	Using intersection surfaces and road section surfaces, we build city blocks (See Fig. \ref{sg.fig:block}).
	We define crudely a city block surface as the complementary surface to its bounding road surfaces and road intersections.
	However, because all the road surface surrounding a city block may not have been generated, we use the road axis instead the road surface as city block limit when road surface is missing.
	
	Because the road axis network has been stored as a topology, getting the surface formed by the road axis surrounding the desired block is immediate. 
	Then, we use Boolean operations to subtract the street and intersection surfaces from the face. This has the advantage that this still provides results when some of the street limiting the block have not been computed, which is often the case in practice.
	By definition, the universal face ("outside") is not used as a city block!
	
	\myimageHL{"illustrations/chap3/streetgen/block/uncomplete_block"}{We generate city blocks by computing the surface that is bounded by associated road surface, road intersection, and road axis when no road surface is available (top of illustration).}{sg.fig:block}{0.95}
	
\subsection{Concurrency and scaling}
	The aim of this work are to model streets for a whole city in a concurrent way (that is several process could be generating the same street at the same time).
	Our choice of method is strongly influenced by those factors, and we use specific design to reach those goals, which are not accessory but essential.
	
	\paragraph{One big query}
	We emphasize that StreetGen is one big SQL query (using various PL/pgSQL and Python functions).\\
	The first advantage it offers is that it is entirely wrapped in one RDBMS transaction.This means that, if for any reason the output does not respect the constraints of the street data model, the result is rolled back (i.e., we come back to a state as if the transaction never happened).
	This offers a strong guarantee on the resulting street model as well as on the state of the input data.
	
	Second, StreetGen uses SQL, which naturally works on sets (intrinsic SQL principle). This means that computing $n$ road surfaces is not computing $n$ times one road surface.
	This is paramount because computing one road surface actually requires using its one-neighbours in the road network graph.
	Thus, computing each road individually duplicates a lot of work.
	
	Third, we benefit from the PostgreSQL advanced query planner, which collects and uses statistics concerning all the tables.
	This means that the same query on a small or big part of the network will not be executed the same way.
	The query planner optimises the execution plan to estimate the most effective one.
	This, along with extensive use of indexes, is the key to making StreetGen work seamlessly on different scales.
	
	\paragraph{One coherent streets model results}
	One of the advantage of working with RDBMSs is the concurrency (the capacity for several users to work with the same data at the same time).\\
	By default, this is true for StreetGen inputs (road network). Several users can simultaneously edit the road axis network with total guarantees on the integrity of the data.
	
	However, we propose more, and exploit the RDBMS capacities so that StreetGen does not return a set of streets, but rather create or update the street modelling.
	\\
	This means that we can use StreetGen on the entire Paris road axis network, and it will create a resulting streets modelling. Using StreetGen for the second time on only one road axis will simply update the parameters of the street model associated with this axis. Thus, we can guarantee at any time that the output street model is coherent and up to date.
	
	Computing the street model for the first time corresponds to using the `insert' SQL statement. When the street model has already been created, we use an `update' SQL statement. In practice, we automatically mix those two statements so that when computing a part of the input road axis network, existing street models are automatically updated and non existing ones are automatically inserted. The short name for this kind of logic (if the result does not exist yet, then insert, else update) is `upsert'.

	This mechanism works flawlessly for one user but is subject to the race condition for several users. We illustrate this problem with this synthetic example. The global streets modelling is empty.
	User1 and User2 both compute the street model $s_i$ corresponding to a road axis $r_i$.
	Now, both users upsert their results into the street table. The race condition creates an error (the same result is inserted twice).
	\myimageHL{"illustrations/chap3/concurrency/upsert_and_race"}{Left, a classical upsert. Right, race condition produces an error.}{sg.fig:upsert}{0.95}
	
	We can solve this race problem with two strategies.
	The first strategy is that when the upsert fails, we retry it until the upsert is successful.
	This strategy offers no theoretical guarantee, even if, in practice, it works well.
	We choose a second strategy, which is based on semaphore, and works by avoiding computing streets that are already being computed.
	
	When using StreetGen on a set of road axes, we use semaphores to tag the road axes that are being processed.
	StreetGen only considers working on road axes that are not already tagged.
	When the computing is finished, StreetGen releases the semaphore. 
	Thus, any other user wanting to compute the same road axis will simply do nothing as long as those streets are already being computed by another StreetGen user.
	This strategy offers theoretically sound guarantees, but uses a lot of memory.

\subsection{Generating basic Traffic information}
	\subsubsection{Introduction} 
	StreetGen is based on tables in a RDBMS. As such, its model is extremely flexible and adaptable.
	We use this capacity to generate basic geometric information needed for traffic simulation.
	The world of traffic simulation is complex, various methods may require widely different data, depending on the method and the scale of the simulation.
	
	%scale
	For instance, a method simulating traffic nation-wide (macro simulation) would not require the same data as a method trying to simulate traffic in a city, neither as a method simulating precise trajectory of vehicle in one intersection.
	
	%semantic
	Moreover, traffic simulation may require semantic data. 
	For instance an ordinary traffic lane and the same lane reserved to bus may be geometrically identical but have a very different impact in the simulation.
	
	%adavanced
	Traffic simulation may require traffic light sequencing, statistics about car speed and density, visibility of objects, lighting, etc.
	
	%what we do 
	Therefore, we do not pretend to provide data for all kind of traffic simulations, but rather to provide basic geometric data at the scale of a city.
	The basic geometric information we choose to provide are lane and lane interconnection.
	Because lane and interconnection are integrated into StreetGen, the links between lane, interconnection and road network (road axis, intersection) is always available if necessary.

	We define lane as the geometric path a vehicle could follow in a road section. A lane is strictly oriented and is to be used one-way.
	The intersections are trajectories a vehicle could follow while in an intersection, to go from one road section (lane) to another road section (lane).
	Similarly, interconnections are one-way.	
	
	\subsubsection{Generating Lanes}
	%geometry
	Our data contains an approximate number of lane per road axis.
	Even in absence of such data, it could be estimated based on the road width and importance.
	
	We compute the lanes of an axis using the buffer operation (formaly Minkowsky sum with disk),
	as a simple translation would not produce correct result (See Fig. \ref{sg.fig:no_translation}).
	We create lane axis and lane separator, the second being a useful representation, and potential base to generate lane separation markings. 
	The lane generation then depends on the parity of the number of lane, and is iterative (See Fig. \ref{sg.fig.lane_from_1_to_6}).
	Special care must be taken so that all polylines generated have a coherent geometric direction.
	 
	\myimage{"illustrations/chap3/lane_parity/lane_from_1_to_6"}{Generating various number of lanes, displayed in QGIS with dotted lines.}{sg.fig.lane_from_1_to_6}
	
	%direction
	Our data set also gives approximate information direction for each road axis.
	The road axis direction may be 'Direct', 'Reverse' or 'Both'.
	'Direct' and 'Reverse' are both for one-way roads, with the global direction being relative to the road axis geometry direction (i.e. order of points).
	In 'Both' case we only know that the road is not one-way.
		 
	Please note that this simple information are very lacking to describe even moderately complex real roads (for instance, 3 lane in one direction, and one lane in the other). 
	For lack of better solution, we have to make strong assumptions.
	
	In the case of 'Reverse' or 'Direct', all lanes shall have the same direction. 
	In the 'Both' case, lanes on the right of the road axis should have same direction as road axis, and lanes on the left opposite direction.
	In odd case, the center lane will be considered on the right of the road axis.
	Lane are numbered by distance to road axis, side (right first).
	Figure \ref{sg.fig.lane_direction} gives an overview of possible lane directions.
	\myimageHL{"illustrations/chap3/lane_direction/lane_direction"}{Default possible direction for lanes.}{sg.fig.lane_direction}{0.95}.

	\subsubsection{Generating trajectories in interconnection}
	%lack of data
	Our dataset lacks any information about lane interconnection, i.e. which connexion between lanes are possible and what trajectory those connections have. 
	For instance, being on the right lane of street X, is it possible to go to the right lane of street Y at tne next intersection, and following which trajectory?
	
	Strong assumptions are necessary.
	We use the orientation of lanes and consider that interconnection can only join lanes having opposite input direction in an intersection.
	Considering an intersection, each lane either comes in or out of this intersection (intersection input direction).
	Furthermore, we consider that lanes of the same road section are not directly connected (no turn around). Please note that in real life usage such trajectory may be possible.
	We create an interconnection for each pair of lanes respecting those conditions.
	
	Actual vehicle trajectories in intersections are very complex, depending both on kinematic parameters, driver perceptual parameters, driver profile, vehicle, weather condition, etc.
	For instance \cite{Wolfermann2011} study a simple case and model only the speed profile.
	
	We generate a plausible and simple trajectory using Bezier curves.
	Moreover, we isolated the part responsible for trajectory computing so it can be easily replaced by a more adapted solution than Bezier curve.
	
	\myimageHL{"illustrations/chap3/interconnection/interconnection"}{interconnection trajectory, Bezier curve influenced by start/end and possibly intersection centre.}{sg.fig.intersection_trajectory}{0.95}
	
	Bezier control points are the points where lane center enter/exit the intersection.
	The third control point depends on the situation. It usually is the barycentre of lanes intersection and intersection centre.
	However, when lanes are parallel, lane intersection is replaced by enter/exit barycentre.
	In special case when lanes are parallel and opposite, the centre of the intersection is not considered to obtain a straight line trajectory.
	Figure \ref{sg.fig.intersection_trajectory} presents interconnection trajectory generation in various situations.
	
\subsection{Roundabout detection}
	\label{sg.method.roundabout}
	 
	StreetGen has been used for traffic simulation.
	StreetGen does not consider semantic difference for any intersection.
	However traffic simulation tools make a strong difference between intersection and round-about. 
	
	Still, the traffic modelling is widely different between roundabout and classical intersection.
	Thus we need a method to detect roundabouts.
	We face a problem similar to (\cite{Touya2010}, Section 3.1).
	The main issue is that round-about definition is based on the driving rules in the intersection (type of priority, no traffic light,...).
	Yet those details are not available on the road axis network we use.
	If we use a strict geometric definition (round about are rounds), we could try to extract the information from aerial images (\cite{Ravanbakhsh2009}) or from vehicle trajectory \cite{Zinoune2012}. Yet both this example are not in street settings,
	where round-abouts may be much smaller, and much harder to see on aerial images.
	Moreover, vehicle trajectory would be much less precise because buildings mask GPS.
	
	Of course we are far from having this level of information, therefore we used the little information available, that is geometrical shape of street axis and street names.
	We need a way to characterize a round-about that can be used for detection.
	We cannot define round-about only based on the topology of the road network (such as : a small loop), nor purely based on geometry (road axis is forming a circle) because round abouts are not necessary round. 
	We noticed that road axis in a roundabout tends to have the same name,
	and/or contain the word 'PL' or 'RPT' (IGN short for 'Place' and 'Rond-Point' (roundabout)).
	
	Therefore we use two criterias to define a potential roundabout : its road axis may be round (geometric criteria), and the road axis might have the same name or contain 'PL' or 'RPT' in their name (toponym criteria).
	We use Hough transform (\cite{Duda1972}) to detect quadruplets of successive points in road axis that are a good support for an arc of circle, then perform unsupervised clustering wia DBSCAN algorithm
	 (\cite{scikit-learn}, \cite{Ester1996}).
	To exploit road-name we explore the road network face by face while considering if all the road of a face have the same name and/or some contains special 'PL' or 'RPT' key words.
	
	The final results are weighted, and are used by an user to quickly detect roundabouts (See Figure \ref{sg.fig.round_about}).
	\myimageHL{"illustrations/chap3/roundabout/roundabout"}{Roundabout detection to reduce an user work.}{sg.fig.round_about}{1}.

\subsection{Street Objects : From Road to Street}
	So far we essentially considered road modelling (road axis, road surface, intersection centre, intersection surface, lane, interconnection, etc.).
	However streets can be characterized by the great number and diversity of objects present in it.
	We use street objects in a broad sense, including street furniture as well as marking, trees, etc.
	Figure \ref{sg.fig.street_objects} gives examples of street objects modelling within StreetGen.
	
	\myimageFullPageWidth{"illustrations/chap3/street_objects/objects_column"}{Street objects in StreetGen. Objects are semantic points with advanced symbology viewed in QGIS. Each object can be positioned and oriented relatively to the street axis or street border.}{sg.fig.street_objects}.
	
	\subsubsection{Generic street objects}
	 
	Streets contains a great number of diverse objects.
	We propose to add those to StreetGen via an extensible mechanism so that the system can easily be completed with complex semantic or hierarchy in the future.
	
	We observe that lot of street objects are related to the street axis and road surface, be it for position or orientation.
	For instance, a pedestrian crossing is defined relatively to the street axis and the sidewalk. Its orientation is also often related to the street axis direction (though this is not always the case).
	
	We then design a system where objects can be linked to street axis, so as to be able to compute orientation and position accordingly.
	Each object position can be defined in absolute coordinates, according to street axis or according to side-walk position. If the object is positioned relatively to street axis or sidewalk, the object position is defined by its the curvilinear abscissa along the street axis.
	\myimageHL{"illustrations/chap3/street_object_principle/street_object_principle"}{Object position and orientation can be defined relatively to road model.}{sg.fig.street_object_principle}{0.95}.
	
	Each object orientation can also be defined as absolute or according to street axis.
	The list of possible objects is defined in a table that can eb easily extended, and also used to build more complex information (for instance, a hierarchy). Triple store format would be good candidates for this.

	Figure \ref{sg.fig.street_object_principle} illustrates the principle of object positioning and orientation relatively to street axis and/or street border.
	Lets take the example of safety barrier. In Paris those are commonly used on sidewalk, few centimeters from roadway, parallel to street axis.
	
	In our current implementation, generic street object underlying representation is a point with semantic and possibly relative positioning and orientation information.
	This generic representation can be specialised for specialised street objects.
	
	\subsubsection{Specialised street objects} 
	
	Generic street objects are numerous in streets, but many objects require more specific parameters and/or different representation than a point.
	We demonstrate that specialised street object can be introduced as specialisation of generic objects.
	The concept is borrowed from inheritance in object programming design.
	A new table needs to be created for each specialized street objects.
	This table is linked to the generic table through foreign key and add-oc triggers.
	The specialised table store additional parameters and additional representation, and is kept synchronised with the relevant objects of the generic table.
	
	We demonstrate this functionality with the object pedestrian crossing 
	(See Fig. \ref{sg.fig.pedestrian_crossing}).

	\myimageHL{"illustrations/chap3/street_objects/pedestrian_crossing"}{Specialized objects can be added, with adequate paremeters (here width) and representation (here surface  displayed with qgis dash pattern).}{sg.fig.pedestrian_crossing}{0.95}

	A pedestrian crossing can be parametrized by its position along the road axis (curvilinear abscissa) and its orientation relative to the road axis. Those two parameters are already defined for generic objects. However, pedestrian crossing also necessitate a width parameter, defining the width of  pedestrian crossing at the road axis level.
	Furthermore, a pedestrian crossing is better not represented by a point and a symbol, but rather by a surface, going from sidewalk to sidewalk, which implies a special geometry (surface rather than point) and a function to generate this surface based on the pedestrian crossing parameters and the road surface.
	
	\myimage{"illustrations/chap3/street_objects/make_parallelogramoid"}{Building the pedestrian crossing surface based on its parameters and road axis and surface.}{sg.fig.make_parallelogramoid}.
	
	Creating the pedestrian crossing surface is not immediate because road axis is a polyline. We create such function by first creating points delimitating the pedestrian crossing on the road axis. 
	Then those points are projected left and right with an angle onto the road surface. 
	Then the road surface border between points is extracted and sewed together to form the parallelogramoid.
	Figure \ref{sg.fig.make_parallelogramoid} illustrates pedestrian crossing surface creation.

%	\newpage
	
%% Inserting the Result part
	%% ---------------------------------------------------------------------
%% Copyright 2014, Thales, IGN, Rémi Cura
%% 
%% This file present the result of the article
%% ---------------------------------------------------------------------

\section{ Results }
	\label{sg.sec:result}
	This section is dedicated to testing our street modelling method.
	Our road model relies on turning radius, as such we start by an experiment about estimating those turning radius.
	We then test the core of StreetGen, with experiments on result quality, robustness, scaling, concurrency and parallelism.
	The we test the traffic information generated in a real world traffic simulation application.
	Last, we test how generic StreetGen is by generating challenging roads, roads in another country, and an airport runway. 
	
\subsection{Estimating default turning radius} 
\label{sg.result.estimating_radius}
	Due to lack of information about the turning radius, we had to make the assumption that turning radius depends on the type of roads (see Section \ref{sg.method.kinematic_hypothesis}).
	Although this hypothesis is a good start and has been empirically verified outside cities,  it has not been tested for city as far as we know.
	\myimageHL{"illustrations/chap3/radius_analysis/radius_analysis_workflow"}{Workflow to test radius hypothesis.}{sg.fig.radius_analysis_workflow}{1}
	
	Sadly radius data is not available for Paris, therefore we propose a framework to estimate this hypothesis for Paris.This test framework is illustrated in figure \ref{sg.fig.radius_analysis_workflow}.
	
	%\todobetter{chap3:result:guessing radius : put pseudo code for reader comfort}
	
	First we extract arc of circles from Open Data Paris 'trottoir' (sidewalk) layer using a kind of Hough transform (\cite{Duda1972}) and filtering.
	Result is noisy because layer 'trottoir' contains many round objects that are not cornerstones.
	We obtain about 14 thousand arcs of circle.
	Then we map a GPS road network database containing approximate driving speed with the BDTopo road network which contains road importance, road width, number of lane, etc.
	We use a fuzzy geometric (how much space is shared by road axis dilated by few meters) and fuzzy semantic distance (comparing both axis street name using trigram).
	See Fig. \ref{sg.fig.radius_and_road_speed}.
	\myimageHL{"illustrations/chap3/radius_analysis/radius_and_road_speed"}{Radius detection and average road speed obtained by various dataset merging.}{sg.fig.radius_and_road_speed}{1}
	
	Based on road data, we try several ways to predict turning radius.
	The first method "guesstimate" is to manually design a simple function 'radius = $f_{guess}$(road importance)'. The second method is to use the results of french SETRA which link average vehicle speed and turning radius for peri-urban roads, with 'radius = $f_{speed}$(average speed of vehicles)', using equation \ref{sg.eq.setra}.
	\begin{equation}
	\label{sg.eq.setra}
	f_{speed}(speed, width)=18.6*\sqrt{\frac{speed}{\lvert10.0*width+65.0-speed\rvert}}
	\end{equation}	
	
	Lastly we use machine learning to train a random forest regressor using road importance, speed, and road width to predict the radius, thus having a 'radius = $f_{rforest}$(importance, speed, road width)'.
	Random forest prediction is intended as a comparison to other two methods.
	Results are given in table \ref{sg.tab.radius_prediction_error} and illustrated in \ref{sg.fig.radius_several_functions}.

	\begin{table}[!htb]
		\centering
		\caption{Error for various radius prediction method}
		\footnotesize
		\label{sg.tab.radius_prediction_error} 
		\begin{tabular}{|c|c|c|c|}
			\hline metric (\metre)&   $f_{guess}$& $f_{speed}$  & $f_{rforest}$ \\ 
			\hline average abs error & 2.41 & 2.18 & 1.97 \\ 
			\hline median abs error & 1.9 & 1.91 & 1.69 \\ 
			\hline 
		\end{tabular} 
	\end{table}
	
	\myimage{"illustrations/chap3/radius_analysis/radius_several_functions"}{Illustrating radius predicted with various methods, for major and residential roads.}{sg.fig.radius_several_functions} 

\subsection{StreetGen} 
In this section we perform tests on StreetGen core.
    \paragraph{Robustness}
	Overall, StreetGen generates the entire Paris road network. 
	We started by generating a few streets, then a few blocks, then the sixth arrondissement of Paris, then a fourth of Paris, then the entire south of Paris, then all of Paris.
	Each time we changed scale, we encountered new special cases and exceptions. Each time we had to robustify StreetGen.
	We think it is a good illustration of the complexity of some real-life streets and also of possible errors in input data.
	
	\paragraph{Quality}
	Overall, most of the Paris streets seem to be adapted to our street data model. 
	StreetGen results looks primarily realistic, even in very complex intersections, or overlapping intersections.
	\myimage{"illustrations/chap3/streetgen/success/success"}{Example of results of increasingly complex intersection.}{sg.fig:success}
	
	Results are un-realistic in a few borderline cases (see Figure \ref{sg.fig:failure}), either because of the hypotheses or the limitations of the method.
	Those cases are, however, easily detected and could be solved individually.
	
	\myimage{"illustrations/chap3/streetgen/failure/failure"}{Various cases of failure from more severe to less severe (1 , 2 , 3). 1 : loop, 2 : bad buffer use, 3 : radius too big for network.}{sg.fig:failure}
	
	Failure 1 is caused by the fact that axis 1 and 2 form a loop. Thus, in some special cases, the whole block is considered an intersection. This is rare and easy to detect. \\
	Failure 2 is caused by our method of computing intersection surface. In a T junction, a large street orthogonal to a small street will produce a bump. It could be dealt with using the variable buffer.\\
	Failure 3 is more subtle and happens when one axis is too short with respect to the radius. In this case, the end of the arc is way out of the intersection, because the intersection is so short. It could be fixed by taking into consideration the next axis with roughly the same direction, but it would introduce special cases.

    We compared the result of StreetGen with the actual roadways of Paris, which are available through Open Data Paris\footnote{\url{http://opendata.paris.fr/page/home/}}. It clearly shows the limit of the input data, chiefly in roadway width estimations. 
    \myimageFullPageWidth{"illustrations/chap3/streetgen/correcting/correcting"}{The estimated parameters may be far from reality (Left). It is, however, possible to manually or automatically fit the street model byy editing the model parameters.}{sg.fig:correcting}
    Using interactive tools (See Fig. \ref{sg.fig:correcting}), it is possible to update the input data so that it is closer to the reality, until a very good match is reached.
    A qualitative evaluation of StreetGen result is not possible without the parameter of the road model being adapted to fit to reality, which is performed manually or automatically in further work.

    \paragraph{Scaling}
    The entire Paris street network is generated in less than 10 minutes (1 core). Using the exact same method, a single street (and its one-neighbour) is generated in $\sim 200 \space \milli \second$, thus is lower than the human interactive limit of $\sim300\space \milli\second$.
    
    \paragraph{SQL set operations}
    We illustrate the specificity of SQL (working on set) by testing two scenarios.
    In the first scenario (no- set), we use StreetGen on the Paris road axis one-by-one, which would take more than $2 \space hours$ to complete.
    In the second scenario (set), we use StreetGen on all the axis at once, which takes about $10\space minutes$.
    
    \paragraph{Concurrency}
    We test StreetGen with two users simultaneously computing two road axis sets sharing between 100\% and 0\% of road axis. The race condition is effectively fixed, and we get the expected result.
    
	\paragraph{Parallelism}
    We divided the Paris road axis network into eight clusters using the K-means algorithm\footnote{\url{http://scikit-learn.org/stable/modules/generated/sklearn.cluster.KMeans.html}} on the road axis centroid (See Fig. \ref{sg.fig:clustering}).
    This happens within the database in a few seconds.
    Then K users use StreetGen to compute one cluster (parallelism), which reduces the overall computing time to about one $minute$.
	
	\myimageHL{"illustrations/chap3/clustering/clustering_road_axis_centroid"}{Clustering road axis centroid with K-Means, K=20, (black segments are convex hull).}{sg.fig:clustering}{1}
	
\subsection{Using Streetgen for traffic simulation}
	In this section we design an experiment to test the usefulness of StreetGen-generated traffic information.
	
	Visually, generated lane and interconnection seems to be adapted in most case.
	The computation cost is however significantly increased because lane and interconnection are generated by triggers, and not at a global level.
	Moreover interconnection uses PLPython and shapely, which introduce a strong overhead.

	We test StreetGen usability for traffic simulation by exporting its model for SymuVia, a traffic simulation tool (\cite{Leclercq2007}). 
	This work was performed by Lionel Atty (IGN, SIDT) for the project TrafiPollu (\cite{Soheilian2016}),
	using a mix of sql query and python modules orchestrated in a QGIS plugin.

	\myimageHL{"illustrations/chap3/to_simuvia/to_simuvia"}{Converting StreetGen Traffic model to SymuVia traffic model.}{sg.fig:to_simuvia}{1}
	
	The principal difficulties were different handling of intersection (SimuVia intersection model require semantic, like roundabout of classic intersection, see Section \ref{sg.method.roundabout}), necessity to regroup lanes having the same direction into homogeneous lane groups (See Fig. \ref{sg.fig:to_simuvia}), simplification of geometries and XML export to custom SymuVia format.
	
	Exporting is not fast (10 $\minute$ for a hundred streets), but exported data is successfully used in the SymuVia traffic simulation tool.

	Figure \ref{sg.fig:comparison_symuvia_sg} proposes an example of road network traffic information manually created in a custom Symuvia tool and the same data automatically created automatically with StreetGen.
	Automated results where sucessfully used in SYmuvia traffic simulation tool, although it outlined the imprecison of StreetGen input data (regarding number of lanes).
	 \myimageFullPageWidth{"illustrations/chap3/to_simuvia/comparison_symuvia_sg"}{Manually created traffic information and StreetGen automatic traffic information.}{sg.fig:comparison_symuvia_sg}

\subsection{Extending Streetgen applications} 
	StreetGen was designed with Paris city in mind, that is many heritage roads. Indeed, street layout and characteristics vary widely around the world, and special knowledge about city type greatly helps creating adapted hypothesis.
	For instance a Manhattan-like grid layout is much easier to deal with. 
	
	We can still test StreetGen genericity and robustness.
	To this effect, we use StreetGen in different unusual scenarios.
	
	\paragraph{intra city fast lane (Lille)}
	\myimage{"illustrations/chap3/extended_usage/lille/lille_sg_stereopolis"}{Experiment on StreetGen genericity : High speed intra city road, Lille, France.}{sg.fig:lille}
	The first scenario is to use StreetGen to model roadway of a modern part of Lille with fast roads
	 (See fig. \ref{sg.fig:lille}).
	We stress that this example does not contains bridge or over passes, as StreetGen cannot manage those.
	Those roads have a modern design and as such do not necessary follow StreetGen hypothesis.	
	StreetGen model was however sufficiently generic to model well the roadway.
	This road model was edited from scratch in one our, using in-base interaction.
	
	\paragraph{Grid-based layout (Mali)}
	\myimage{"illustrations/chap3/extended_usage/mali/mali"}{Experiment on StreetGen genericity : Mostly grid-based city layout, Mali}{sg.fig:mali}
	The second scenario is to use StreetGen for a grid-based road network (See Fig. \ref{sg.fig:mali}).
	We do not have access to ground truth for this area.
	However, the results proved to be satisfactory for further use in 3D world building.
	
	\paragraph{Airport (Bergen)}
	\myimage{"illustrations/chap3/extended_usage/airport/airport_sg_gmap"}{StreetGen used on airport runway and service roads, Bergen, Norway}{sg.fig:airport}
	The last scenario is more stretched, and proposes to use StreetGen to generate airport runway and servicing roads (See Fig. \ref{sg.fig:airport}).
	With very few exception StreetGen is able to model the runaway surface (and servicing roads) of the airport. 
	
	However because all dimensions are so different from Paris street (runway width of 40 meter is not uncommon in airport), this scenario higlighted the need to be able to change default settings easily.
	To cater for this need, we changed all StreetGen settings to be stored in a global settings table.
	Then this glbal settings table can be adapted to each situation.
	
%	\newpage

%% Inserting the Result part
	%% ---------------------------------------------------------------------
%% Copyright 2014, Thales, IGN, Rémi Cura
%% 
%% This file is the Discussion of the result of the article
%% ---------------------------------------------------------------------

 \section{ Discussion }
	\label{sg.sec:discussion}
	
	\subsection{Estimating default turning radius}
	In Section \ref{sg.result.estimating_radius} we experimented to evaluate how the turning radius could be estimated.
	
	Results in Table \ref{sg.tab.radius_prediction_error} are pretty un-conclusive. 
	Whatever the function to predict radius, results are poor. 
	Radius extracted are simply too noisy, and the average speed information we extracted from road speed database lacks details (only 4 values possible).
	Even a random forest regressor could not work properly.
	Intuitively, the radius probably depends also on road construction date, historical data, neighbourhood, or other data.
	The SETRA function $f_{speed}$ (\ref{sg.eq.setra}) results are however relevant when dealing with major roads (See fig. \ref{sg.fig.radius_several_functions}), but not in general.
	
	We tested the hypothesis that maybe the function was correct but was badly parametrised.
	To this end, we tried to find the optimal parameters for $f_{speed}$ using non linear least square optimisation with loss function to reduce outliers weight.
	We could not find better values.
	We take that as a proof that we do not have sufficient data to conclude about this function overall fitness for our need.

	This experiment would be better performed using an adjusted StreetGen results.
	
	\subsection{Street data model}
	Our street data model is simple and represents the roadway well, but would need to be detailed in some aspects.\\
	First, parking places are very abundant and important in Paris street layout, yet our model cannot specifically deal with these.\\
	Lanes cannot have different width nor type (bus lanes, bicycle lanes, etc.).
	\\
	Our model is just the first step towards modelling streets.
	Because we model streets, our model can not deal with bridge, tunnels, overpasses,etc.
	This limitations steems from the tools we use for topology management : PostGIS Topology.

	\subsection{Kinetic hypothesis}
	Overall, kinetic hypotheses provide realistic looking results, but are far from being true in an old city like Paris. Indeed, a great number of streets pre-date the invention of cars.
	We attempted to find a correlation between real-world corner radius (analysing OpenDataParis through Hough arc of circle detection) and the type of road or the road's average speed (from a GPS database). 
	We could not find a clear correlation, except for fast roads. 
	On those roads, the average speed is higher, and they have been designed for vehicles following classical engeneering rules.

	\subsection{Precision issue}
	All our geometrical operations (buffer, boolean operations, distances, etc.) rely on PostGIS (thus GEOS\footnote{\url{http://trac.osgeo.org/geos/}}).
	We then face computing precision issues, especially when dealing with arcs. Arc type is a data type that is not always supported, and thus it must be approximated by segments.
	\myimage{"illustrations/chap3/streetgen/precision_issue/precision_issue"}{Example of a precision issue. Left, we approximate arcs with segments, which introduces errors. Right, the error was sufficient to incorrectly union the intersection surface.}{sg.fig:precision_issue}
	
	StreetGen uses various strategies to try to work around these issues.
	However the only real solution would be to use an exact computation tool like CGAL \cite{cgal:eb-15a}. It would also allow us to compute the circle centres in 3D.
	A recent plugin called SFCGAL\footnote{\url{www.sfcgal.org/}} integrates parts of CGAL in PostGIS, using exact computing
	
	\subsection{Streetgen for traffic} 
	Export is successful but a bit slow, although slowness is largely due to the simuvia XML format.
	One of the principal problem is that too many interconnections are generated.
	Indeed, we generate all possible interconnection, we would greatly benefit from an heuristic to generate only plausible interconnections.
	
	\subsection{Street objects} 
	Street objects addition to StreetGen greatly improves modelling possibilities.
	The system was however not tested at full scale.
	At the city scale, storing all objects in one table may prove to be a limitation (Paris contains dozen of millions of street objects).
	We demonstrated point based and surface based objects, however line-based objects like markings are also very prominent.
	
	The main limitation is however that we did not test automatic object generation based on rules and patterns.
	We indeed consider that correctly solving the object problem requires grammars or similar high level semantic tools which we did not try.
	
	\subsection{Extend use for StreetGen}
	Using a tool slightly oustide of its intended functionality is always interesting.
	Among the limitation, the Lille roadway possessed one road with linearly increasing road width, which can not be modelled by StreetGen.
	Mali dataset revealed a problem when input road axis are not properly topological.
	Model an airport is clearly a stretch of StreetGen capabilities.
	In particular, airport runways posses lots of semantic objects like lights, beacons, etc.
	The difference is obvious when comparing StreetGen results with an actual airport modelling (Fig. \ref{sg.fig.thales_airport}, courtesy of Thales TTS).
	\myimageFullPageWidth{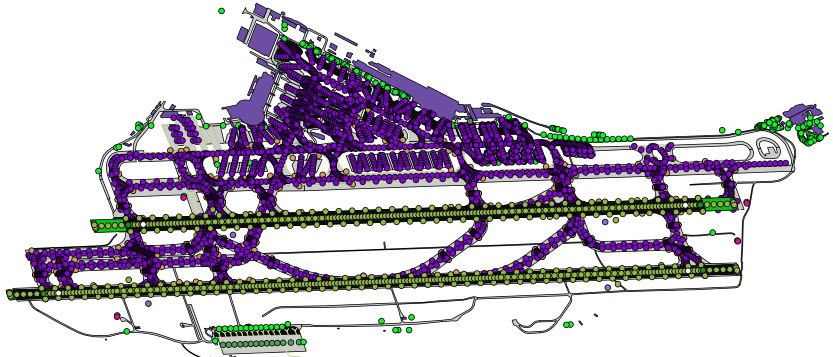}{Real airport model, courtesy of Thales TTS}{sg.fig.thales_airport}
	
	\subsection{Fitting street model to reality}
		
		StreetGen was designed from the beginning to provide a best guess of streets based on very little information.
		However, in some cases, we want the results to better fit reality.\\ 
		For this, we created an interactive behaviour so that several users can fit the automatic StreetGen results to better match reality (using aerial images as ground truth for instance).
		
		We did not created a Graphical User interface (GUI), but rather a set of automatic in-base behaviours so that editing input data or special interaction layers can interactively change the StreetGen results. Doing so ensures that any GIS software that can read and write PostGIS vector can be used as StreetGen GUI.
		\\
		In some cases, we may have observations of street objects or sidewalks available, possibly automatically extracted from aerial images or Lidar, and thus imprecise and containing errors.
		We tested an optimisation algorithm that distorts the street model from best-guess StreetGen to better match these observations.
		
		This subject is similar to Inverse Procedural Modeling, and we feel it offers many opportunities.

%	\newpage

%% Inserting the conclusion
	%% ---------------------------------------------------------------------
%% Copyright 2014, Thales, IGN, Rémi Cura
%% 
%% This file is the conclusion of the article
%% ---------------------------------------------------------------------

\section{Conclusion} 
As a conclusion, we proposed a relatively simple street model based on a few hypotheses. This street data model seems to be adapted to a city as complex as Paris. We proposed various strategies to use this model robustly. 
We showed that the RDBMS offers interesting possibilities, in addition to storing data and facilities for concurrency.

Our method StreetGen has ample room for improvements. We could use more sophisticated methods to predict the radius, better deal with special cases, and extend the data model to better use lanes and add complex street objects managed by grammars.

In our future work, we also would like to exploit the possibility of the interaction of StreetGen to perform massive collaborative editing.
Such completed street modelling could be used as ground truth for the next step, which would be an automatic method based on detections of observations like side-walks, markings, etc. Finding the optimal parameters would then involve performing Inverse Procedural Modelling. 

\section{Acknowledgment}
This article is an extract of \cite{Cura2016thesis} (chap. 3). We thank Prof.Peter Van Oosterom and Prof.Christian Heipke for their extensive review.
%	\newpage

% %Note : to generate up to date biblio: go to zotero, select all library, the right click and export as better bibtex, save a cura2014 in the root folder.

	\section{Bibliography}  
	\bibliography{./all_bibli} 

%%%%
%% Inserting the todos if in proofreading mode
%	\begin{proofreading}
%	
%	%%% Inserting the annexe, contains supp. mat. that must be choosed
%	%	\input{./src/annexe.tex}
%	%	\newpage
%		
%
%	%% Inserting the todo list of remaining stuff
%	 	% %manual general todo (top down)
%	 	\newpage
%	 	\input{./src/todo.tex}
%	 	% %automatic precise todo (bottom up)
%		\newpage
%		\todototoc
%		\listoftodos
%		
%	\end{proofreading}

\end{document}